\documentclass[3p,preprint,12pt]{elsarticle}

\usepackage{amssymb}
\usepackage{amsmath}
\usepackage{color}
\usepackage[dvipsnames]{xcolor}
\usepackage{multirow}
\usepackage{caption}
\usepackage{subcaption}
 
\usepackage{hyperref}
\usepackage{cleveref}
\usepackage{mathtools,leftindex,tensor,mhchem}
\usepackage{graphicx, color, transparent}
\usepackage{setspace}
\usepackage{bm}
\usepackage{euscript}
\usepackage[normalem]{ulem}

\begin{document}

\begin{frontmatter}

\title{Inverse design of programmable shape-morphing kirigami structures}

\author[1]{Xiaoyuan Ying}
\author[1]{Dilum Fernando}
\author[1]{Marcelo A. Dias\corref{cor1}}
\ead{marcelo.dias@ed.ac.uk}
\cortext[cor1]{Corresponding author}
\address[1]{Institute for Infrastructure \& Environment, School of Engineering, The University of Edinburgh, Edinburgh~EH9~3FG Scotland, UK}

\begin{abstract}
Shape-morphing structures have the capability to transform from one state to another, making them highly valuable in engineering applications. In this study, it is propose a two-stage shape-morphing framework inspired by kirigami structures to design structures that can deploy from a compacted state to a prescribed state under certain mechanical stimuli---although the framework may also be extended to accommodate various physical fields, such as magnetic, thermal, and electric fields. The framework establishes a connection between the geometry and mechanics of kirigami structures. The proposed approach combines the finite element analysis (FEA), genetic algorithm (GA), and an analytical energy-based model to obtain kirigami designs with robustness and efficiency.
\end{abstract}

\begin{keyword}
Kirigami \sep Shape-morphing structures\sep Mechanical metamaterials \sep Optimisation\sep Inverse design
\end{keyword}

\end{frontmatter}


\section{Introduction}

Kirigami is a traditional Japanese art of paper cutting that has evolved into a powerful technique used by scientists attempting to address various engineering problems related to deployable, stretchable, and flexible structures. The fundamental principle of kirigami design is to incorporate functionalities into a flat and thin sheet through strategically placed cuts, voids, and perforations. By designing and arranging a cut pattern, the local mechanics of the fundamental motifs~\cite{sadik2021local,sadik2022local} endow a compact thin sheet with the ability to transform into desired 2D and 3D shapes through a coordinated motion of rigid panels and flexible ligaments. This concept has inspired engineering applications in various fields, including metamaterials~\cite{bertoldi2017flexible,yang2018multistable,ali2022metamaterials,cardoso2021structural}, soft robotics~\cite{cheng2020kirigami,jin2021mechanical,he2023modular,cui2018origami}, flexible actuators and electronics~\cite{dias2017kirigami,sun2018kirigami,won2019stretchable,yang2023new,li2022breathable}, and energy harvester~\cite{qi2022kirigami,qu2020kirigami}.

Kirigami tessellations~\cite{grunbaum1987tilings}, for example, formed by a series of cuts and panels, have unique material properties such as super-stretchability and auxeticity, which are desirable mechanical properties. More interestingly, the deformation path and deployed shape can be predetermined by tuning the sizes, orientations, and topology. This gives raise to an intriguing problem: whether it is possible to create a general framework that produces a cut pattern that, if subjected to external stimuli and/or boundary conditions, deploys into a prescribed shape. To address this problem, it is crucial to conduct a comprehensive analysis of the kinematics and mechanics of kirigami tessellations. The kinematics refers to the motion and deformation of the kirigami tessellation, where the panels can be seen as rigid body and all panels are connected by perfect rotational hinges with no friction. While the mechanical aspect refers to the forces, stresses, and energy involved in the process and how the deployment governed by stimuli and boundary conditions. The forward analysis of kirigami tessellation was explored in prior research ~\cite{grima2000auxetic,andrew2004negative,yin2022notch,rafsanjani2017buckling,zheng2023modelling,czajkowski2022conformal}, where the geometric laws and principles of kirigami tessellation were established. In the context of forward analysis, it is referred to studies that are conducted once the design of the kirigami tessellation has been determined. However, the majority of research focuses on regular and periodic cutting patterns, where the kinematics of a tessellation behaves as a single degree-of-freedom system. This allows for analytical derivation of the kinematic equations, which reduces system complexity but limits flexibility in shape-morphing. For inverse design methods, the sensitivity study of periodic square kirigami tessellation was conducted to investigate how hinge width and gap width affect the in-plane and out-of-plane behaviour~\cite{yin2022notch,lamoureux2015dynamic}. For general kirigami tessellations, an inverse design framework, with sufficient geometry constraints for non-rigid deployable kirigami tessellations, was proposed for creating non-periodic cutting patterns that allowed for more complex deployed shapes~\cite{choi2019programming}. However, such study did not have insights in the mechanics and it was agnostic of the deployment path. In the meanwhile, a theorem for rigid deployable kirigami tessellation was proposed in 2D~\cite{dang2021theorem} and 3D~\cite{dang2022theorem} scenarios, and the deployment path was understood. Furthermore, an additive framework was proposed that could produce a reconfigurable and rigid-deployable cut pattern~\cite{dudte2023additive}, and the deployment path of kirigami tessellation could be derived analytically. This framework provides an insight to generate shape-morphing kirigami tessellation. Based on the framework, a physically-aware differentiable design of magnetically actuated kirigami was proposed \cite{wang2023physics} in order to realise target shape-morphing under specific magnetic excitation. However, this framework can only work for homogeneous loading conditions where the deployment paths are determined based on the kinematic principle and every panel can rotate freely to the prescribed position. In the case of inhomogeneous boundary conditions, where the deployment path is highly related to the applied excitation, the inverse design framework remains uncovered. 

\begin{figure*}
    \centering
    \includegraphics[width=1\textwidth]{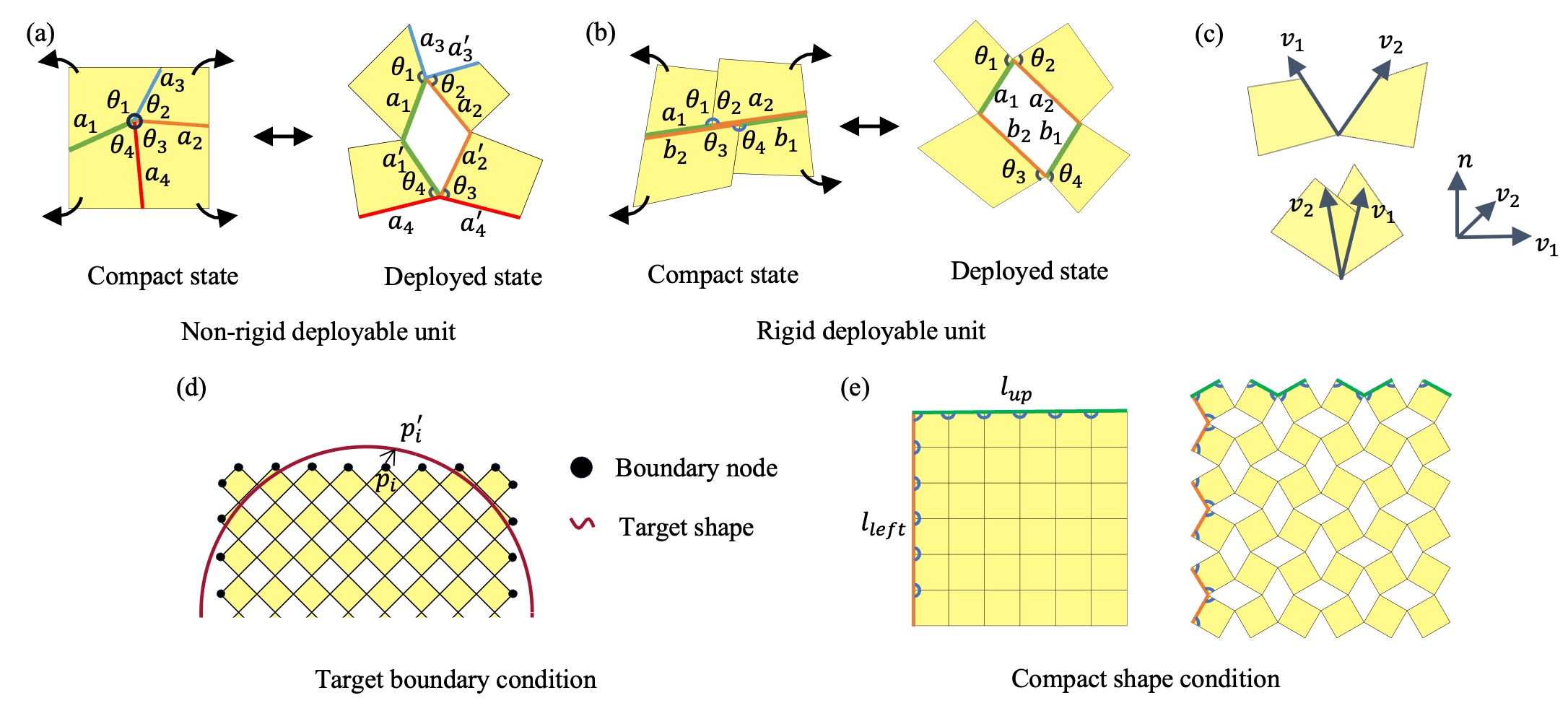}
    \caption{Geometric constraints: (a) Edge and angle conditions for non-rigid deployable units which allowing the units close seamlessly. (b) Edge and angle conditions for rigid-deployable units which allowing the unites close seamlessly. (c) Non-overlapping deployment, using outer product to prevent adjacent units from overlapping. (d) Target boundary condition. (e) Compact shape condition, allowing the compacted state to be a certain configuration.}
    \label{geometric constraints}
\end{figure*}

In the context of engineering applications, inhomogeneous conditions are a far more common occurrence than ideal homogeneous conditions. The solution to overcome this challenge will have a profound impact in real world situations. In this paper, a discussion will be given of the inverse design of kirigami tessellation under inhomogeneous boundary conditions in order to address the existing knowledge gap.
In general, inverse design for shape-morphing structures combines two tasks: (1) a forward method that can accurately predict the deformed shape under certain external stimuli, and (2) an inverse optimisation framework that can efficiently find the acceptable solution to achieve the target shape. For the first task, several methods have been implemented to address periodic Kirigami tessellation, purely rotating model using classic beam theory~\cite{rafsanjani2017buckling}, coarse-grained model based on kirigami effective (cell-averaged) deformation ~\cite{czajkowski2022conformal,zheng2023modelling}, hybrid mechanism model considering ligaments as hybrid system of bending springs, shear springs and longitudinal springs~\cite{coulais2018characteristic}. However, for non-periodic kirigami tessellations, the above methods are difficult to capture the accurate deformation. Here, FEA is used to predict the deformed shape taking into account the geometric and material nonlinearity. For the second task, evolutionary algorithm (EA) can be used as a technique to find satisfactory solutions to global optimisation problems. For example, Particle swarm optimisation (PSO) was adopted to find the best structure design by tuning parameters~\cite{perez2007particle}. Genetic algorithm (GA) was applied to design the topology of cuts that can lead to prescribed shape~\cite{xue2017kirigami}. Machine learning (ML) was used to design shape-morphing soft smart beam combined with neural network (NN)~\cite{ma2024machine}. 

This study proposes a two-step optimisation framework for shape-morphing kirigami structures. The first step utilises a constrained kinematic optimisation to generate the cut pattern that allows the kirigami tessellation to deploy into the target shape while satisfying geometric compatibility. In the second step, the generated cut pattern is used as the initial cut guess and is further optimised considering elasticity. This produces a cut pattern that can be stretched into a target shape under certain boundary condition. The proposed method reduces the computational cost and bridges the kinematics and mechanics of kirigami tessellations of varying complexity.

\section{Methodology}

\subsection{Kinematics}

The basic deformation of a rigid kirigami tessellation can be described as a series of quad units rotating around the out-of-plane axis, while the axis translates due to dilation. This transformation can be denoted simply by matrix representation. For two dimensions, the three transformations, namely rotation, translation, and scaling, are written. The rotation matrix $\mathbf{R} (\xi )$ is given by 
\begin{equation}
    \centering
    \mathbf{R}(\xi ) = \begin{bmatrix}
    \centering
    \cos\xi  & -\sin\xi & 0\\
    \sin\xi  & \cos\xi & 0\\
    0 &  0&0
    \end{bmatrix},
    \label{eq:1}
\end{equation}
where $\xi$ is the rotating angle. 
The translation matrix $\mathbf{T}$ is determined by 
\begin{equation}
\centering
\mathbf{T} = \begin{bmatrix}
\centering
1  & 0 & t_{x}\\
0  & 1 & t_{y}\\
0 &  0&1
\end{bmatrix},
\end{equation}
where  $t_{x}$ and $t_{y}$ are the displacements in x and y directions, respectively.
Finally, the scaling matrix $\mathbf{S}$ is determined by following equation, 
\begin{equation}
\mathbf{S} = \begin{bmatrix}
\centering
s_{x} & 0 & 0\\
0  & s_{y} & 0\\
0 &  0& 0
\end{bmatrix},
\end{equation}
where $s_{x}$ and  $s_{y}$ are the scaling factors in x and y directions respectively.

Furthermore, a concise overview will be given of two types of kirigami tessellation with different geometric constraints based on previous study~\cite{choi2019programming,dang2021theorem,dudte2023additive}. The first condition is non-rigid deployable kirigami tessellation with one intersection. The second condition is a rigid-deployable kirigami tessellation with two intersections. Rigid-deployability refers to the ability to deploy the tessellation by rotating the panels around the corners while maintaining their size, angle, and topology. On the contrary, for a non-rigid deployable kirigami tessellation, the geometric conditions for its compact and deployed states are guaranteed, however, geometric incompatibility could happen during the deployment. 

\begin{figure*}[h!]
    \centering
    \includegraphics[width=1\textwidth]{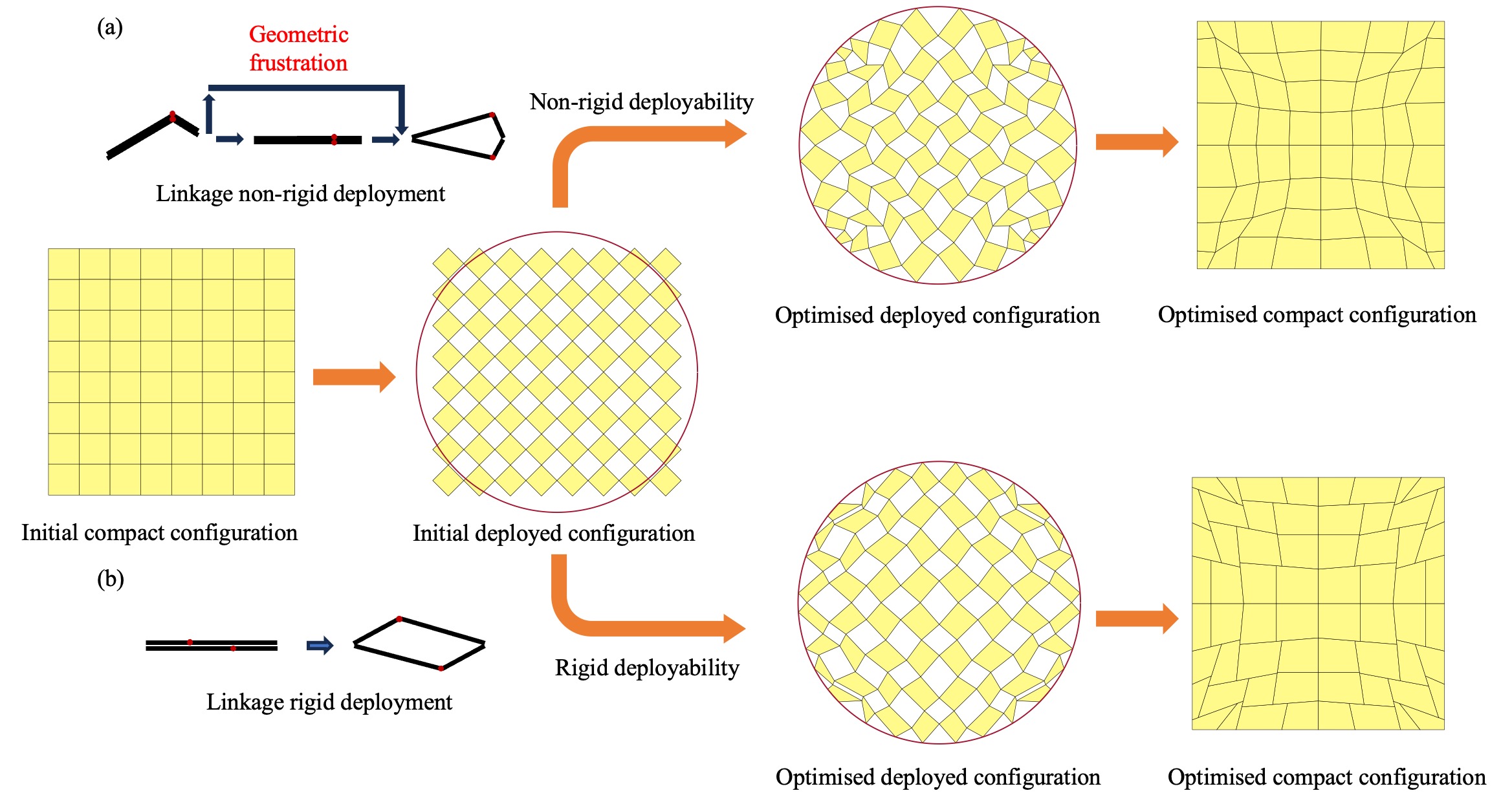}
    \caption{Kinematic optimisation. (a) Non-rigid deployable optimisation. (b) Rigid deployable optimisation. The figure shows the process of kinematic optimisation of kirigami tessellation from initial compact configuration, using different geometric constraints can achieve different optimised configuration.  The initial deployed configuration is taken as initial input, and using non-rigid deployability condition or rigid-deployability condition to get the corresponding optimised configurations, and use a simple kinematic derivation to acquire the compact configurations.}
    \label{kinematic optimisation}
\end{figure*}

\subsubsection{Non-rigid deployable tessellation}
Here, quad kirigami tessellations are considered. 
To ensure that the kirigami tessellation is composed of quad units without voids, the contractility and compatibility constraints should be satisfied, as shown in \cref{geometric constraints}. More explicitly, these conditions are stated as follows:
\begin{enumerate}
    \item {\it Edge condition}. The length of every corresponding edges should be the same (see \cref{geometric constraints}(a)):
\begin{equation}
        a_{i} ^{2}-a_{i} ^{\prime\,2}=0.
\end{equation}
\item {\it Angle condition}. The sum of every angles around an interior corner equal to 2$\pi$ (see \cref{geometric constraints}(a)):
\begin{equation}
    \sum \theta _{i} =2\pi.
\end{equation}
\item {\it Non-overlapping condition}. During the deformation planar should be prevented from overlapping, which can be ensured by using following constraint (see \cref{geometric constraints}(c)):
\begin{equation}
    \left \langle \vec{v_{1}} \times \vec{v_{2}},\hat{n}   \right \rangle \ge 0,
\end{equation}
where $\hat{n} = (0,0,1)$ is the outward unit vector, and $\vec{v_{1}}$ and $\vec{v_{2}}$ are the vector of adjacent edges forming a right-handed order.
\item {\it Boundary shape condition}. The boundary shape of deployed kirigami tessellation should match the target shape. The difference between deployed shape and target shape can be describe by the projection distance from deployed boundary nodes to the boundary shape (see \cref{geometric constraints}(d)):
\begin{equation}
    \left |  \right | p_{i}-\tilde{p} _{i}\left |  \right |^{2} =0,
\end{equation}
where $p_{i}$ is boundary nodes and $\tilde{p} _{i}$ is the projection of $p_{i}$ onto the prescribed boundary shape.
\item {\it Compact shape condition}. In order to get the specific compact configuration, for an example a square, following conditions should be satisfied (see \cref{geometric constraints}(e)):
\begin{equation}
        l_{\mathrm{left}} -l_{\mathrm{up}} =0.
\end{equation}
\end{enumerate}

The kinematic optimisation process can be seen in \cref{kinematic optimisation} for non-rigid deployability optimisation.

\subsubsection{Rigid-deployable tessellation}
For rigid-deployable tessellation, it has been demonstrated~\cite{dang2021theorem} that quad kirigami tessellation is rigidly deployable if, and only if, all cuts form parallelogram voids. Thus, the following conditions must be satisfied: 
\begin{enumerate}
    \item {\it Edge condition}. The opposite edges should be of equal length, thus forming a parallelogram void (see \cref{geometric constraints}(b)):
\begin{equation}
        a_{i} ^{2}-b_{i} ^{2}=0
\end{equation}
\item {\it Angle condition}. To ensure rigid-deployability, the sum of angles around every corner should be $\pi$ (see \cref{geometric constraints}(b)):
\begin{equation}
    \theta_{1}+\theta_{2}=
    \theta_{3}+\theta_{4}=\pi
\end{equation}
\end{enumerate}

Similarly, for rigid-deployable kirigami tessellation the non-overlapping condition, boundary shape condition, and compact shape condition should also be satisfied.

The kinematic optimisation process can be see in \cref{kinematic optimisation} rigid-deployability optimisation. The program starts with a compact square kirigami tessellation and, subsequently, utilise simple kinematic laws to employ the deployed square kirigami tessellation as an initial guess. Then the non-rigid deployable conditions and rigid deployable conditions are adopted separately to identify the kinematic configurations that can be reached by the deployed state on the target shape.

\begin{figure*}
    \centering
    \includegraphics[width=1\textwidth]{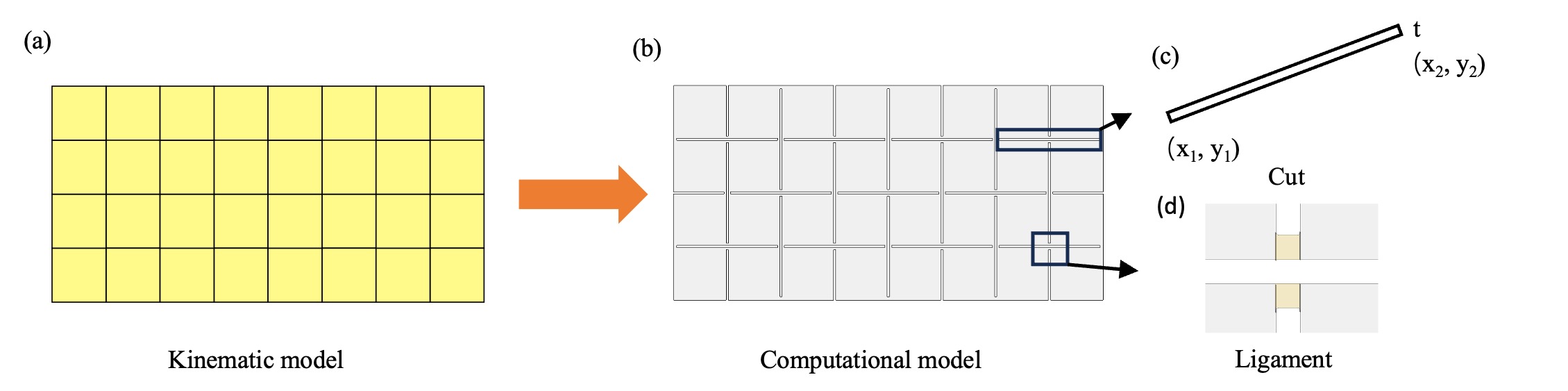}
    \caption{Kirigami tessellation modelling. (a) Kinematic model of kirigami tessellation, the space between each panel is infinitely small. (b) Computational model, each panels are connected by thin ligaments with width and length. (c) Cuts region, each cut is straight line and can be represented by the coordinates of points, width and thickness of cuts. (d) Thin ligaments connecting panels, which can be seen as combinations of bending spring, longitudinal spring and shear spring.}
    \label{Computational modelling}
\end{figure*}

\subsection{Mechanics}

\begin{figure}[ht]
    \centering
    \includegraphics[width=0.46\textwidth]{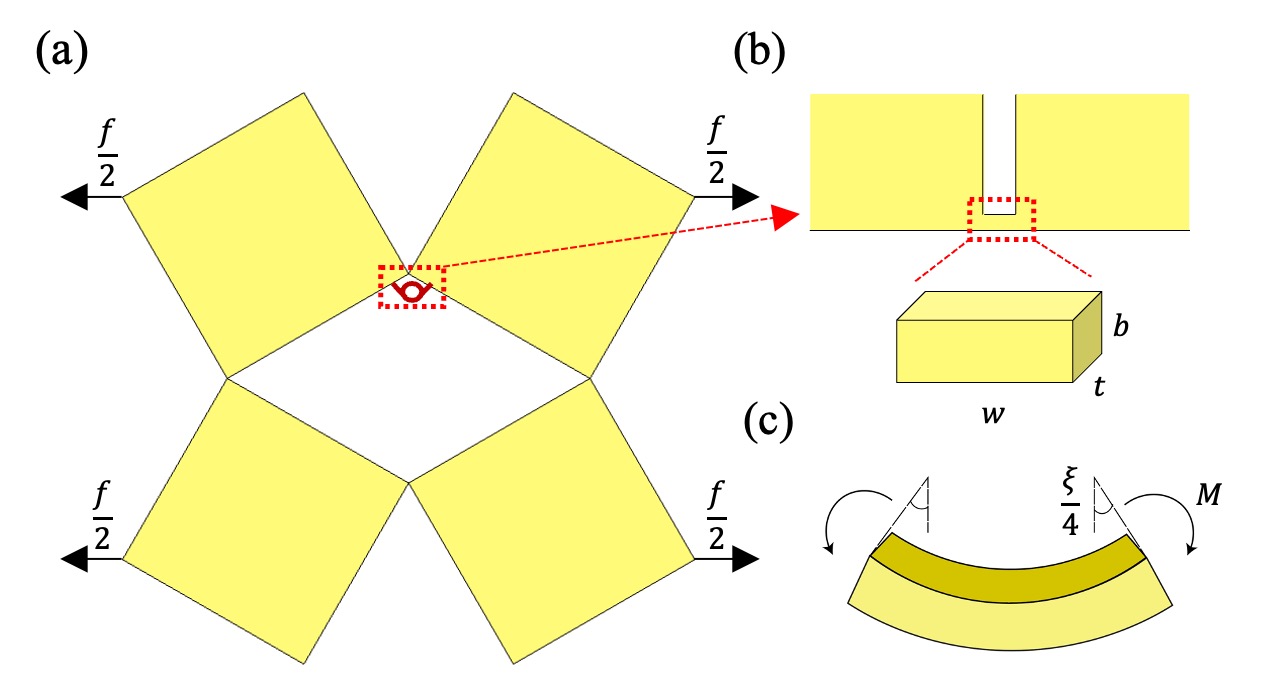}
    \caption{Schematic of rotating spring model: (a) Schematic of deployed tessellation unit. (b) Schematic of spring model of tessellation ligaments. (c) Schematic of spring under bending deformation. }
    \label{rotational hinge model}
\end{figure}

FEA was adopted to get an accurate description of deformed shape and how the deployment process is influenced by mechanics. The simulations were performed using static analysis solver. The neo-Hookean method is used to capture the hyperelastic constitutive behaviour of rubber sheet, and geometric nonlinearities were taken into consideration. The set-up of the computational model is illustrated in \cref{Computational modelling}. From the analytical viewpoint, the kirigami tessellation is modelled as rigid panels connected by thin ligaments, where it is assumed that all deformations are localised at the ligament parts, while the square domain is considered as rigid body (see \cref{rotational hinge model}). Using classical beam theory, it is possible to derive an analytical expression for bending energy\cite{rafsanjani2017buckling}. For simplicity, it is assumed that the each ligament can be seen as a rotational spring that is solely subjected to bending moment, and the rotation of each ligament is penalised by elastic energy, which is determined by bending stiffness $K_b$. This can be written as the following equation:
\begin{equation}
U = \frac{1}{2} \sum_{i=1}^{n}K_b\xi_i ^{2},  
\end{equation}
where $U$ is the total elastic energy of kirigami tessellation, $n$ is the number of ligaments, and $\xi _{i}$  is the corresponding opening angle.

\begin{figure*}[h!]
    \centering
    \includegraphics[width=0.9\textwidth]{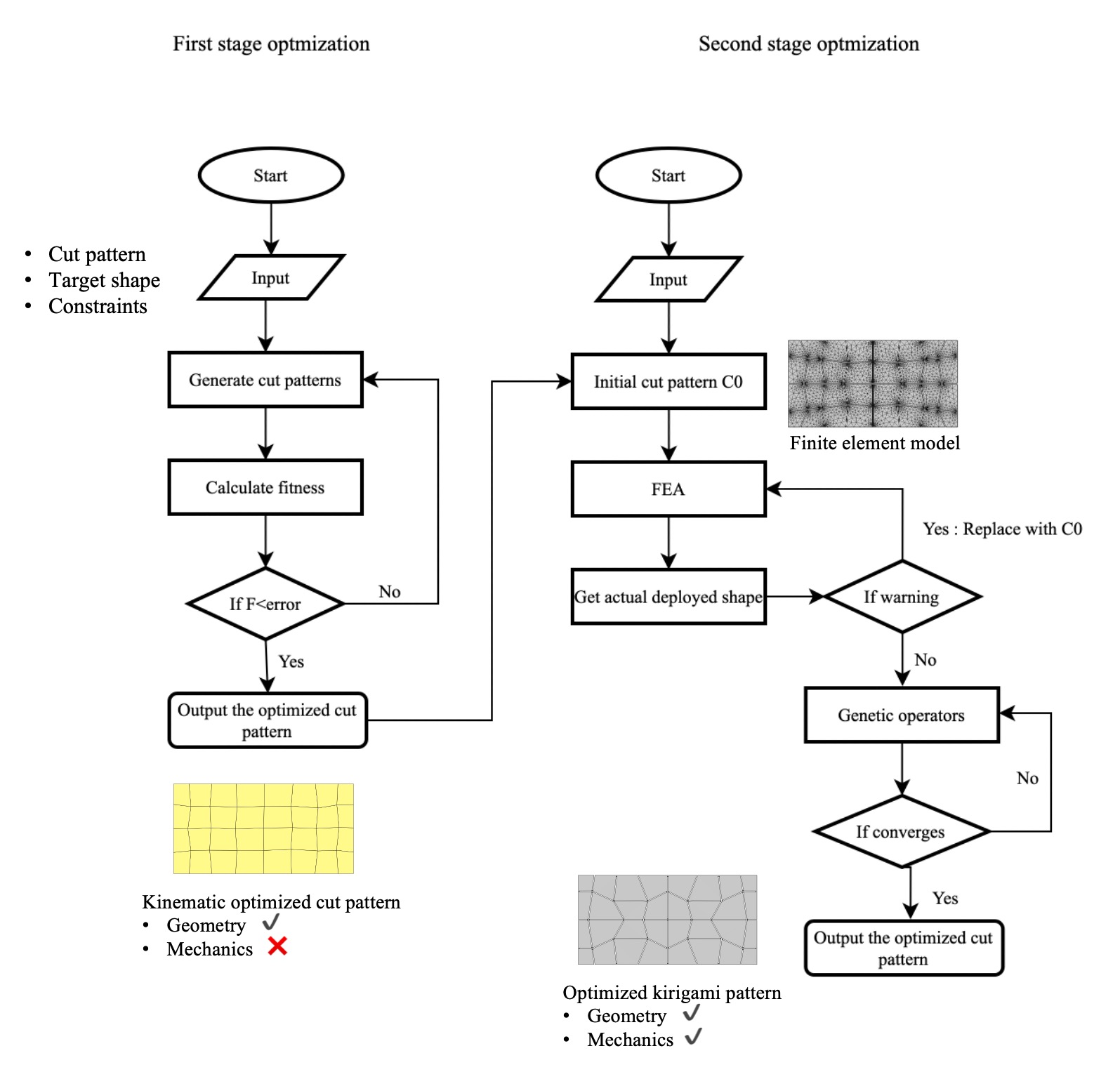}
    \caption{The framework of FEA-GA-based two-step inverse design comprises the following steps: firstly, kinematic optimisation is achieved through the use of a gradient-based algorithm, which takes into account the geometric constraints, target shape and initial guess. The kinematic valid cut pattern is generated in the first step. The generated cut pattern is imported as an initial guess in the second step, using GA to identify feasible cut patterns and using FEA to determine the actual deployed shape under specified boundary conditions (in this instance, forces are applied to the nodes on the shorter sides). This process enables the identification of both mechanically and kinematically valid kirigami patterns.}
    \label{optimisation framework}
\end{figure*}

Here, a single unit is taken for analysis and it is possible to derive an analytical expression for effective bending stiffness $K_b$. As shown in \cref{rotational hinge model}, a square unit deforming uniaxially under the force $f$ is tracked by the degree-of-freedom given by an opening angle $\xi _{i}$. The whole process can be seen as quasi-static.  For simplicity, it is assumed that the loading direction is uniaxial throughout the entire process.
The applied uniaxial stress generates an identical bending moment $M$ at each hinge. This results in the induced rotation of all rigid squares by an opening angle $\xi _{i}$. For each hinge, it is easy to calculate that:
\begin{equation}
M = \frac{EI_{i} }{\rho_{i} },
\label{eq.12}
\end{equation}
where $I_{i} = b^{3}t/12$ is the second moment area of ligaments, and $\rho_{i}$ is the curvature of hinge. Therefore, it may also written that
\begin{equation}
M = \frac{Eb^{3}t\xi _{i} }{12w}. 
\end{equation}
The hinge can be seen as a beam of thickness $t$, width $b$ and length $l$. From using the beam theory, the energy of a ligament can be calculated:
\begin{equation}
U_\mathrm{ligament} = \frac{1}{2}  \int_{0}^{l } \frac{M^{2} }{EI}dx =\frac{Etb^{3}\xi ^{2}  }{24w }.
\end{equation}

\subsection{Two steps of inverse design framework}
To generate a mechanical-aware programmable shape-morphing kirigami tessellation that can deploy onto prescribed shape under certain boundary condition, the kinematics and mechanics of Kirigami tessellation are combined using a GA and FEA. GA is used to generate multiple kinematically valid cut patterns and use scripted modelling to transform kinematic models into finite element models and obtain the actual deployed shape $w(x)$. It is assumed that each cut $C_{i}$ can be represented by a thin straight void region, to simplify this problem, it is assumed that each ligament has the same width $b$ and the same thickness $t$, so each cut can be parameterised by vertices $(x_{1} ,y_{1} )$ and $(x_{2} ,y_{2} )$ (see \cref{Computational modelling} (c)).
The objective function minimises the difference between actual deployed shape and the target shape, which can be expressed as
\begin{equation}
        I = \int_{L}^{} (w(x)-w(\bar{x} ))^{2} ds,
\end{equation}
where $w(\bar{x} )$ is the target boundary shape and $L$ is the boundary region. Hence, the optimisation framework for finding mechanical-aware cut pattern can be formulated as follows:
\begin{equation}
\left\{\begin{matrix}
\mbox{Find:}\quad \ &C = (C_{1} ,C_{2} ,...,C_{n})^{T}  \\\mbox{Minimise:}\quad \ &I = \int_{L}^{} (w(x,C)-w(\bar{x} ))^{2} ds \\ \mbox{s.t.:}\quad \ &C_{i}-r \le C_{i}\le C_{i}+r,
\end{matrix}\right.
\label{optimization framework}
\end{equation}
where $C_{i} = [x_{i1},y_{i1},x_{i2},y_{i2}]$, $n$ is the number of cuts, $w(x,C)$ is the corresponding boundary shape with cut pattern $C$ under tensile displacement, and $r$ is bounds of the cut.

Because of material and geometric nonlinearities, FEA is used to obtain the actual mechanical response. Despite the higher computational cost of this method, an improvement is achieved by using the result generated by kinematic optimisation~\cite{choi2019programming,dang2021theorem} as initial configuration of the joint optimisation as well as by reducing the variables from symmetry conditions. GA is adopted to find the optimal solution of \cref{optimization framework}, as it can solve the considered problem in a robust manner. In order to enhance the efficiency of GA, the mutation rates are adjusted in accordance with the assessment outcomes of the respective offspring in the subsequent generation \cite{thierens2002adaptive}. The mutation rate is defined as a range of values between 0.01 and 0.1, based on the best fitness of a one generation. In order to avoid the issue of non-convergence of the FEA, which could result in an interruption of the optimisation procedure, the {\it try-catch block} (matlab statement) is used to catch errors and replace them with the best individual from the previous generation, thus maintain progress in the optimisation process. To reduce the computational cost, only 1/4 of the structure is modelled, considering the geometry and boundary condition symmetry. The optimisation framework is depicted by the flow-chart in \cref{optimisation framework}.

\subsection{Experimental design}
\begin{figure}[ht]
    \centering
    \includegraphics[width=0.48\textwidth]{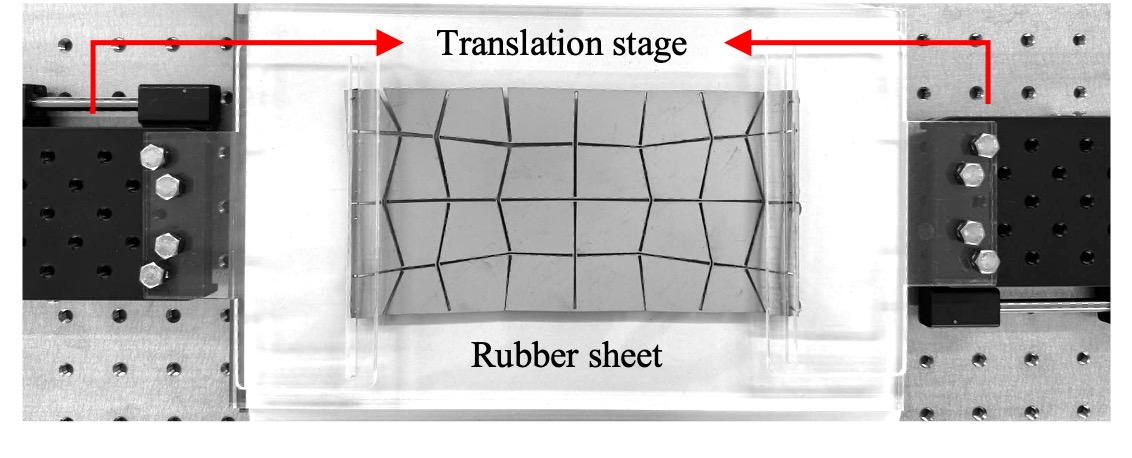}
    \caption{Experiment setup. The rubber sheet is positioned on the smooth glass surface and connected to the translation stage with fixtures, with the deformation being captured by a camera. }
    \label{test setup}
\end{figure}

The samples are fabricated with rubber sheets using a laser cutter. All experiments were conducted on a smooth glass surface to reduce friction and to prevent out-of-plane buckling. The deformation of the rubber sheet was controlled using manual linear translation stages. The setup can be seen in \cref{test setup}. The rubber sheet mounted on a fixture with translation stage allowing rubber sheet stretch in the horizontal direction.

\section{Results and discussion}
This section studies non-rigid and rigid deployable conditions separately. By providing two examples, concave, convex, it is demonstrated the effectiveness of the optimisation framework in each situation. Additionally, comparisons between each case were made in terms of energy. Experiments were conducted based on the optimisation results to verify FEA accuracy. The kirigami tessellation is parametrically modelled.

\begin{figure}[ht]
    \centering
    \includegraphics[width=0.48\textwidth]{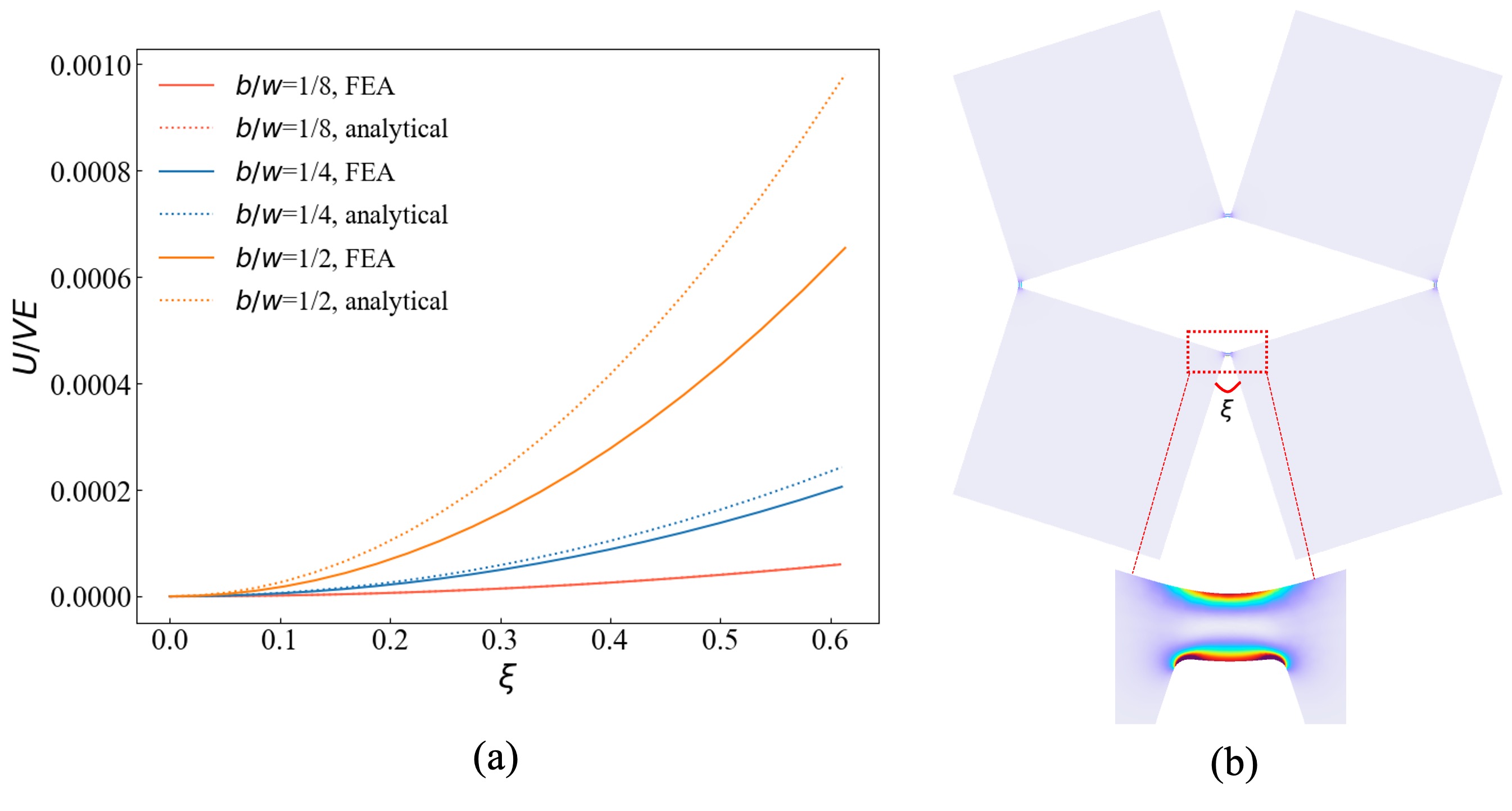}
    \caption{Sensitivity study:(a) A comparative analysis of the elastic hinge elastic energy during deployment between the FEA and analytical model regarding different b/w. [U/VE] is nondimensional quantity, U is the elastic energy of ligaments, V is the volume of ligaments and E is young's modulus. (b) elastic energy density of deployed kirigami tessellation unit}
    \label{fig:sensitivity study}
\end{figure}

\begin{figure*}[h!]
    \centering
    \includegraphics[width=0.77\textwidth]{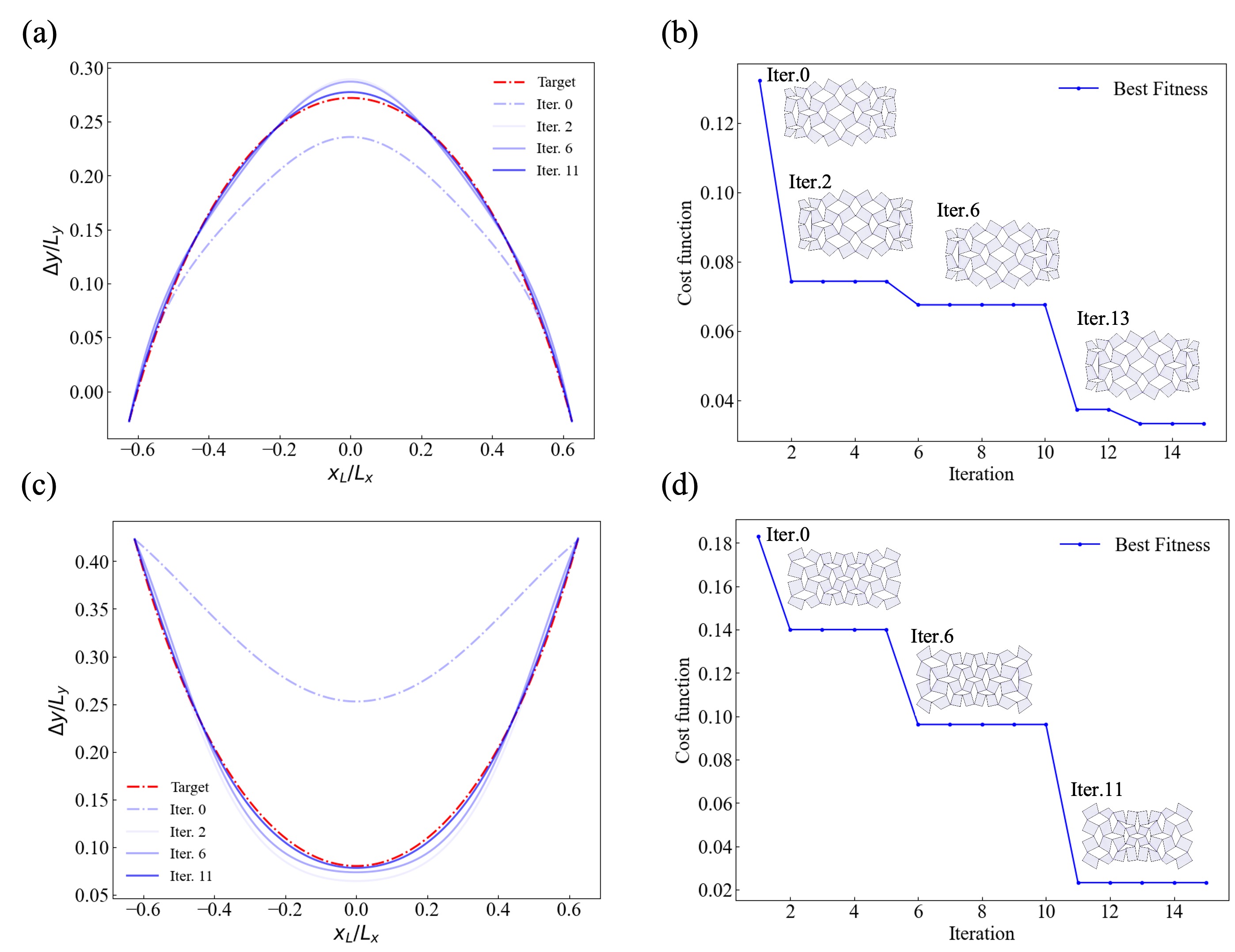}
    \caption{Non-rigid optimisation results: (a) Boundary shape during iterations(convex case), red dotted line is the target shape and ``iter'' stands for the iteration numbers during optimisation process and iter.0 is the initial boundary shape. (b) Evolution curve (convex case), the objective functions during iterations and the actual deployed shape in some iterations. (c) Boundary shape during iterations (concave case). (d) Evolution curve (concave case).}
    \label{non-rigid optimisation}
\end{figure*}

\begin{figure*}[h!]
    \centering
    \includegraphics[width=0.77\textwidth]{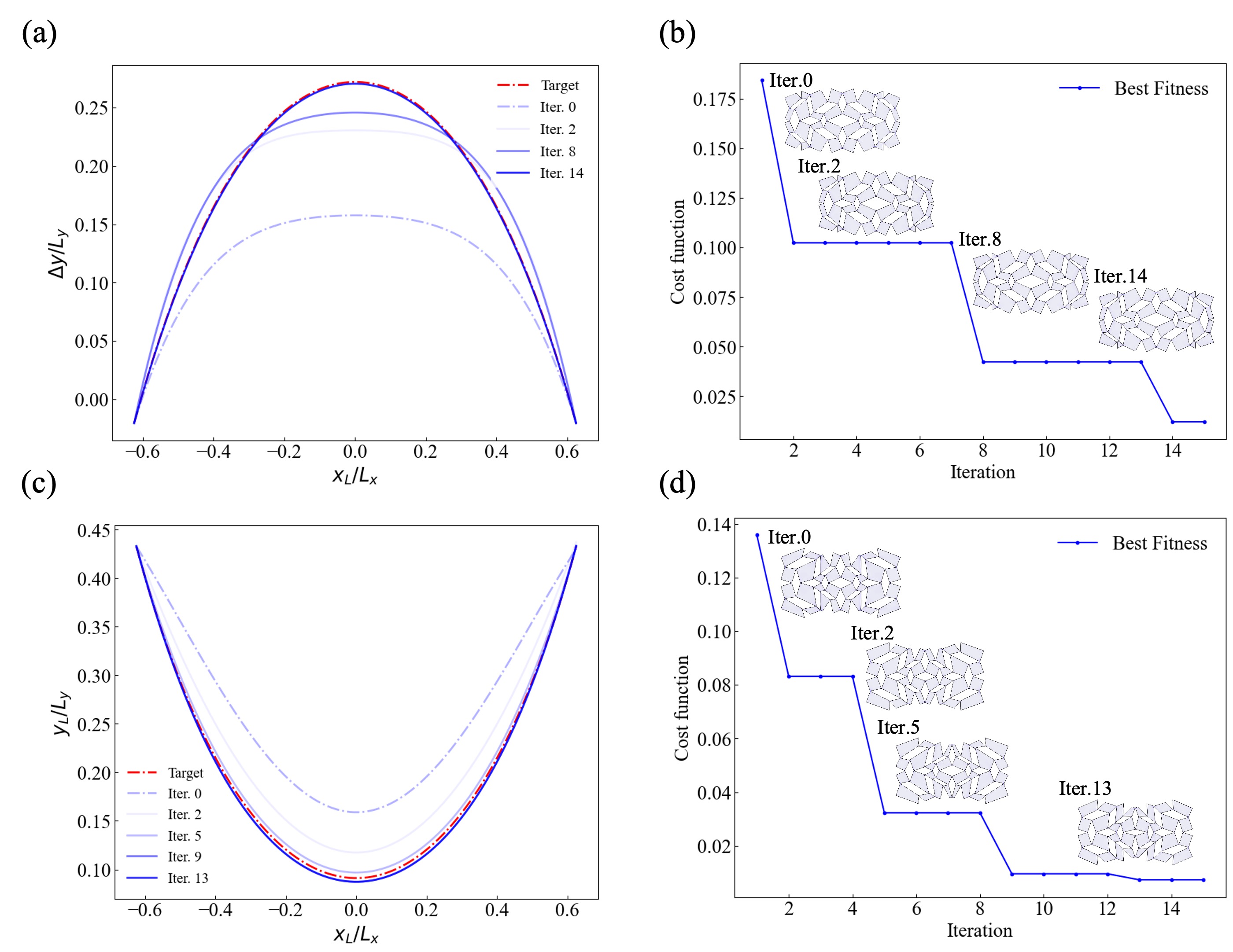}
    \caption{Rigid optimisation results: (a) Boundary shape during iterations(convex case). (b) Evolution curve(convex case). (c) Boundary shape during iterations(concave case). (d) Evolution curve(concave case).}
    \label{rigid optimisation}
\end{figure*}

A sensitivity study was conducted firstly to determine the parameters of model regarding the height $b$ and length $w$ of the ligaments using FEA and proposed analytical model. Three cases were considered with $b/w = 1/8$, $b/w = 1/4$, $b/w = 1/2$, respectively. The results are presented in \cref{fig:sensitivity study}.  In \cref{fig:sensitivity study}(a), it is demonstrated that as the length-to-height ratio $b/w$ increases, the ligament energy also increases, which is consistent with the analytical model. It can be observed that the FEA result is in agreement with the analytical result for low $b/w$ values and only for small opening angles. However, there are significant discrepancies between the two results for large $b/w$ values and large opening angles. This is due to the fact that the Euler-Bernoulli beam theory is limited to conditions where the beam is slender and the deflections are small. As can be observed elastic strain energy density diagram~\cref{fig:sensitivity study}(b), the majority of the elastic strain energy is concentrated in the ligament region. Additionally, the edge of the plate connected to the ligaments also experiences a small amount of elastic energy. Considering the sensitivity study and manufacturing difficulty, the length and height of ligaments are predetermined to be 0.4 mm and 0.2 mm, respectively, and the thickness of rubber sheet was selected to be 3 mm.

\subsection{Optimisation}
Firstly, the non-rigid deployable condition is considered. The results of convex and concave case are presented in \cref{non-rigid optimisation}. It can be observed in \cref{non-rigid optimisation}(a) and (c). The initial boundary shape fails to reach the target shape, whereas the deployed boundary shape reaches towards the target shape during the optimisation process. And the corresponding evolution curve can be seen in \cref{non-rigid optimisation}(b) and (d). It can be observed that the cost function has decreased from 0.15 to 0.02 in convex case and from 0.18 to 0.02 in concave case by following the optimisation process, thus, resulting in a reduction in the discrepancy between the actual deployed shape and the target shape.

Then, the rigid deployability condition is considered. The results of convex and concave case are showed in \cref{rigid optimisation}. It can be observed in \cref{rigid optimisation}(a) and (c). The initial boundary shape fails to reach the target shape at first but gets closer to the prescribe shape after optimisation. The corresponding evolution curve can be seen in \cref{rigid optimisation}(b) and (d). It can be observed that the cost function has decreased from 0.18 to 0.01 in convex case and from about 0.14 to 0.005 in concave case. The optimisation program works both for non-rigid can rigid case and can achieve different target shapes.

\begin{figure*}
    \centering
    \includegraphics[width=1\textwidth]{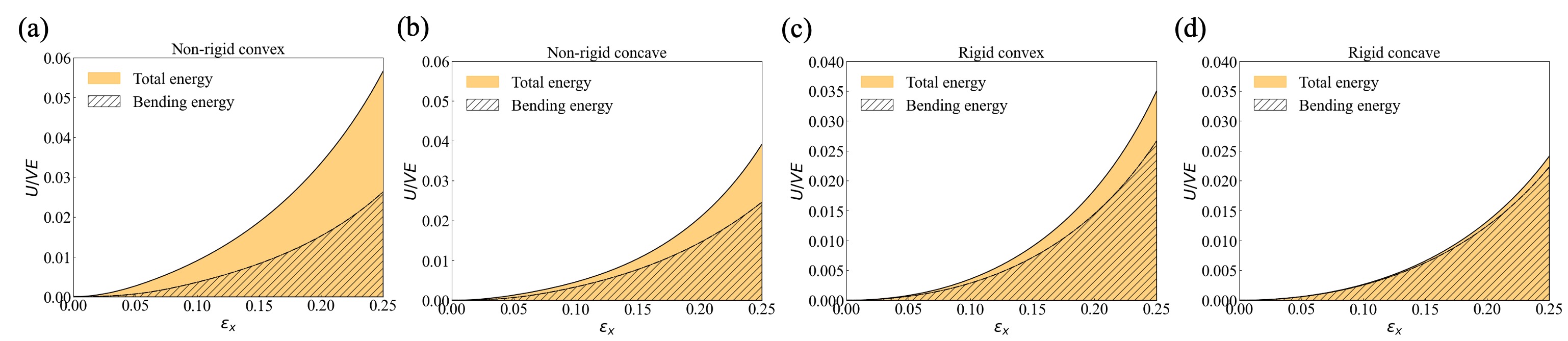}
    \caption{Evolution of total energy and analytically predicted bending energy during deployment for (a) Non-rigid convex case. (b) Non-rigid concave case. (c) Rigid convex case. (d) Rigid concave case.}
    \label{Elastic energy}
\end{figure*}

\begin{figure*}[h!]
    \centering
    \includegraphics[width=1\textwidth]{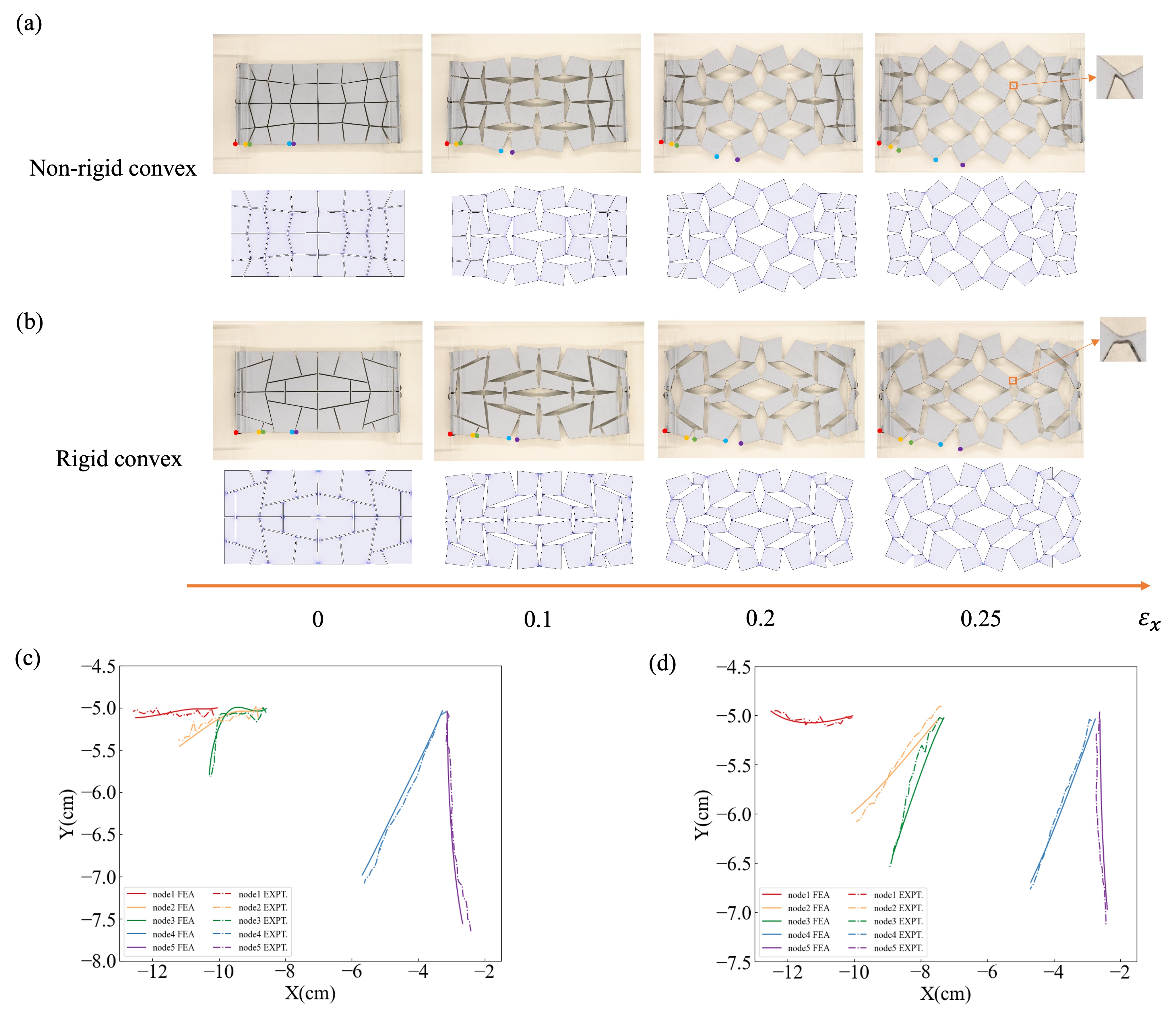}
    \caption{Experiment of (a) non-rigid convex case, the ligaments experience significant shear force. (b) rigid convex case, the ligaments are mostly subjected to bending forces. The deployment shape is presented in terms of test results and simulation results by controlling the horizontal displacement. The deployed shape is presented for $\varepsilon _{x}$ = 0, 0.1, 0.2, 0.25 separately. (c) trajectory of nodes during morphing in non-rigid convex case. (d) trajectory of nodes during morphing in rigid convex case.}
    \label{test results}
\end{figure*}

\begin{figure*}[h!]
    \centering
    \includegraphics[width=1\textwidth]{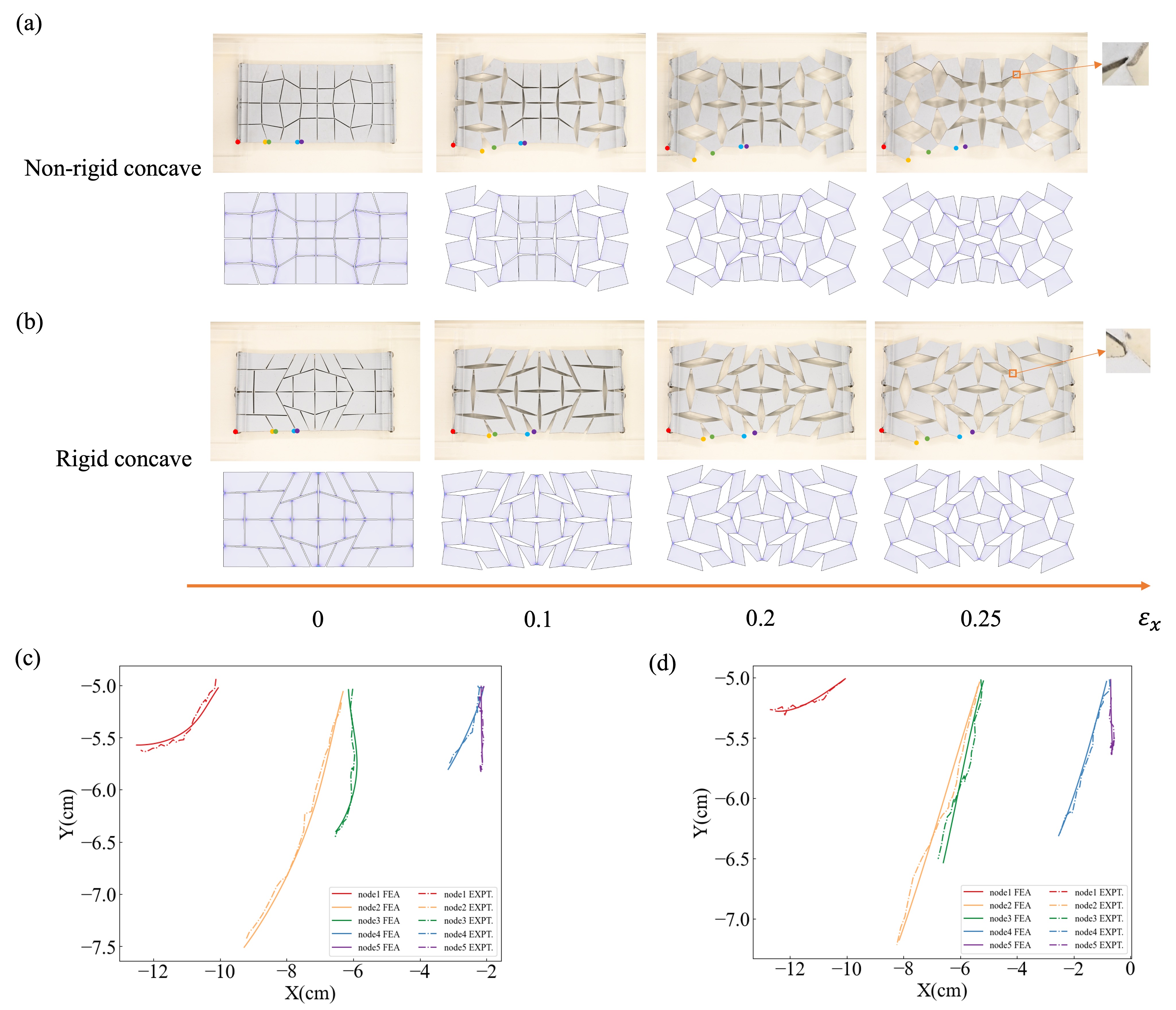}
    \caption{Experiment of (a) non-rigid concave case. the ligaments experience significant shear force. (b) rigid concave case, the ligaments are mostly subjected to bending forces. The deployment shape is presented in terms of test results and simulation results by controlling the horizontal displacement. The deployed shape is presented for $\varepsilon _{x}$ = 0, 0.1, 0.2, 0.25 separately. (c) trajectory of nodes during morphing in non-rigid concave case. (d) trajectory of nodes during morphing in rigid concave case.}
    \label{test results concave}
\end{figure*}

A study was conducted further to investigate the energy evolution during the morphing of optimised kirigami tessellations. The total elastic energy can be obtained from FEA, while the analytically predicted bending energy can be calculated based on the method proposed in section 2.2, see \cref{Elastic energy}. Comparing the non-rigid case \cref{Elastic energy}(a) and (b) and rigid case \cref{Elastic energy}(c) and (d), apparently the rigid deployable tessellations require less elastic energy to deform to the target shape than the non-rigid deployable tessellations. Furthermore, the ratio of bending energy to total energy $U_\mathrm{bending}/U_\mathrm{total}$ in the non-rigid case is less than that in the rigid case with 46\%, 63\% in the non-rigid case and 76\%, 92\% in the rigid case, respectively, indicating that non-bending energy (shear energy and strain energy) is more prevalent in the non-rigid deployment.

\subsection{Experiments}
The experiment setup is introduced in the section 2.4. The kirigami tessellation specimens are fabricated according to the optimisation results with length of 20 cm and width of 10 cm. By controlling the displacement of nodes in horizontal direction, the state of deployment is altered and it realises the sequence of shapes shown in \cref{test results}. For both non-rigid and rigid convex conditions, states are shown at strain values of 0, 0.1, 0.2, and 0.25 in the $x$-direction, denoted as $\varepsilon_x$. It can be seen in \cref{test results}(a) and (b) that the deployed shape has a very good qualitative agreement with the FEA results. As shown in the insets of \cref{test results}(a) and (b), the ligaments in the non-rigid convex case experience significant shear forces, whereas those in the rigid convex condition are predominantly subjected to bending forces. This difference arises because the non-rigid case does not function solely as a mechanism with only the rotational degree-of-freedom provided by the hinge. To provide a more quantitative comparison between simulations and experiments, the path of nodes during the deployment is recorded. This results are given by trajectories of the boundary nodes, as shown in \cref{test results}(c) and (d). The path of nodes in the experiments are consistent with the FEA results-noticed that this process is reversible. Similar results can be observed in the concave cases, as demonstrated in \cref{test results concave}(a), the ligaments in the non-rigid concave case have great shear forces---it is noteworthy that the kirigami tessellation experiences some out-of-plane buckling under large deformation, which originates from local stress relief~\cite{dias2017kirigami,kaspersen2019lifting}. On the other hand, in the rigid concave case, the kirigami tessellation can deploy smoothly in plane, see \cref{test results concave}(b). 

\section{Conclusions}
This study proposes a framework that combines kinematics and mechanics to design programmable shape-morphing kirigami structures that can achieve the prescribed shape when subjected to external stimuli and boundary conditions. In the first stage, sufficient conditions for kinematic optimisation regarding rigid deployable kirigami tessellation and non-rigid deployable kirigami tessellation are put forward to design kinematic valid cut patterns. The analytical model was proposed based on classical beam theory, combined with a sensitivity study to determine the influence of the size of ligaments on elastic energy and further determine the geometric parameters of kirigami tessellations.

In the second stage, GA is used to generate and optimise cut patterns and FEA calculations provide the actual deployed shapes by taking into consideration the mechanics of these systems. To enhance the robustness and efficiency of GA, a cut pattern generated in the initial stage is used as an initial guess in the second stage. Additionally, an adaptive mutation rate is adopted to avoid the optimisation process from becoming trapped in a local minimum. In order to test the efficacy of the optimisation framework, four different cases are considered: non-rigid convex case, non-rigid concave case, rigid convex case, and rigid concave case. Therefore, the optimisation framework presented in this work is capable of producing a kirigami tessellation that can be accurately deployed on the target boundary shape. Furthermore, a comparative analysis was conducted of non-rigid deployable kirigami tessellation and rigid deployable kirigami tessellation with respect to their total energy and energy attribution. Based on the results this optimisation process, experiments were designed and compared to the actual deployed shape with the results generated by FEA during the deployment process.

This research can be extended to facilitate the design of shape-morphing kirigami structures and the presented two-stage framework can accommodate various types of stimuli, including mechanical, magnetic, thermal, electric, and so forth. With an aim to accelerate the optimisation process, future research can be conducted to study the analytical energy-based model of generalised kirigami tessellation relying on hybrid energy and coarse-grained methods~\cite{coulais2018characteristic,czajkowski2022conformal,peng2024programming,zheng2023modelling}. This framework also has the potential to be lifted to 3D structures~\cite{hong2022boundary,konakovic2018rapid}, though relaxation of constraints and adopting more complex boundary conditions.

\section*{Declaration of Competing Interest}
The authors declare that they have no known competing financial interests or personal relationships that could have appeared to influence the work reported in this paper.

\section*{CRediT authorship contribution statement}
MAD acquired funds. XY and MAD designed the research, proposed the methodology, performed the technical implementation, and analysed the results. XY, DF, and MAD wrote the paper.

\section*{Acknowledgements} 
MAD acknowledges UKRI for support under the EPSRC Open Fellowship scheme (Project No. EP/W019450/1).

\bibliography{Bib.bib}

\end{document}